\documentclass[reprint,aps,pre,twocolumn]{revtex4-1}
\usepackage{amsmath,amssymb,mathrsfs}
\usepackage{amsbsy,amsthm,amsfonts}
\usepackage{graphicx}

\begin{document}

\title{Reciprocity relation for the vector radiative transport equation and its application to diffuse optical tomography with polarized light}

\author{Ugo Tricoli, Callum M. Macdonald, Anabela Da Silva, Vadim A. Markel}

\affiliation{Aix Marseille Universit\'{e}, CNRS, Centrale Marseille, Institut Fresnel UMR 7249, 13013 Marseille, France}
\email{ugo.tricoli@frresnel.fr}

\begin{abstract}
We derive a reciprocity relation for vector radiative transport equation (vRTE) that describes propagation of polarized light in multiple-scattering media. We then show how this result, together with translational invariance of a plane-parallel sample, can be used to compute efficiently the sensitivity kernel of diffuse optical tomography (DOT) by Monte Carlo simulations. Numerical examples of polarization-selective sensitivity kernels thus computed are given.
\end{abstract}

\maketitle

Diffuse optical tomography (DOT) employs near-infrared light to probe the macroscopic optical properties of multiply scattering media such as biological soft tissues~\cite{boas_01_1,arridge_09_1}. The typical quantities of interest are the absorption and the scattering coefficients. The inverse problem of DOT is known to be severely ill-posed. Any additional degrees of freedom in the measurements that can alleviate the ill-posedness are of interest. One such degree of freedom is polarization. While the majority of DOT setups employ unpolarized illumination and polarization-insensitive measurements, interest in using polarization has existed since the advent of DOT~\cite{schmitt_92_1,demos_96_1,bartel_00_1}. In particular, it has been demonstrated experimentally that depth sensitivity of the DOT measurements can be improved by using polarization-sensitive measurements~\cite{silva_12_1,rehn_13_1}. Of course, it is understood that strong multiple scattering causes depolarization. Yet, polarization-sensitive measurements can be applied in the mesoscopic scattering regime, that is, on the scale of one or few transport mean free paths $\ell^*$. In biomedical imaging of soft tissues, this translates to physical scales of one to a few millimeters. Typically, sufficiently small source-detector separations and sufficiently short photon trajectories that are still compatible with polarization selectivity are achieved in the backscattering geometry, and we will consider this case below.

To perform DOT reconstructions with polarized light one needs a suitable forward model. In the mesoscopic scattering regime, the commonly accepted mathematical description is based on the vector radiative transport equation (vRTE)~\cite{soloviev_12_1}. One can use vRTE to construct the {\em sensitivity kernel} for DOT. This kernel (defined below in more detail) quantifies the variations of the measured signal due to medium heterogeneities. The linearized inverse problem of DOT can be solved by standard methods once the sensitivity kernel has been computed. The sensitivity kernel for the diffusion equation can easily be defined analytically~\cite{markel_04_4}. However, generalization of this result to the scalar (unpolarized) transport equation has been obtained only recently~\cite{schotland_07_1,machida_16_1} and is of considerable mathematical complexity. A similar analytical result for vRTE is presently not available. Instead, the contemporary mainstream approach to solving vRTE is to use Monte Carlo (MC) simulations~\cite{ramella-roman_05_1,doronin_14_1}. However, application of MC simulation to computing the sensitivity function can be so time-consuming as to render the approach impractical. In this Letter, we derive a reciprocity relation for the vRTE Green's function (a generalization of the known reciprocity relation for the scalar RTE) and show that it can be used to reduce the computational load dramatically. Then we show examples of computed sensitivity kernel for various states of incident and detected polarization.

\begin{figure}
\begin{center}
\includegraphics[width=65mm]{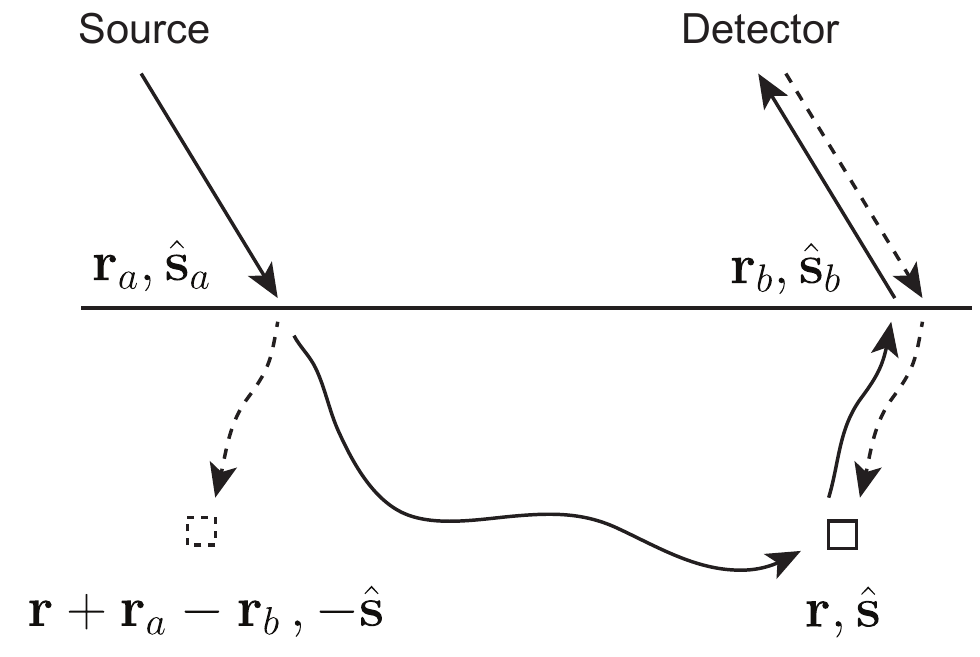}
\caption{Backscattering imaging geometry and illustration of various geometrical objects that are relevant to the reciprocity principle that is considered in this Letter.}
\label{fig:coords}
\end{center}
\end{figure} 

We start with a description of the typical DOT setup in the backscattering geometry (Fig.~\ref{fig:coords}). A single continuous-wave and collimated laser beam is incident at some location ${\bf r}_a$ and in the direction of the unit vector $\hat{\bf s}_a$ on the planar surface of a multiply-scattering medium. A detector measures the intensity of light exiting on the same side of the medium at a different point ${\bf r}_b$ and in the direction $\hat{\bf s}_b$. Inside the medium, the specific intensity ${\mathcal I}({\bf r}, {\bf s})$ obeys the vector radiative transport equation (vRTE)~\cite{ishimaru_84_1,soloviev_12_1}
\begin{align}
( \hat{\bf s} \cdot \nabla + \mu_t) {\mathcal I}({\bf r}, \hat{\bf s}) = \mu_s \int Z (\hat{\bf s}, \hat{\bf s}^\prime) {\mathcal I}({\bf r}, \hat{\bf s}^\prime) d^2s^\prime + {\mathcal S}({\bf r}, \hat{\bf s}) \ .
\label{eqn:rte}
\end{align}
\noindent
Here  ${\mathcal I} = (I,Q,U,V)$ is a vector of the four Stokes components, ${\mathcal S}$ is the source term, $Z$ is the $4\times 4$ phase matrix, $\mu_t$ and $\mu_s$ are the total extinction and the scattering coefficients of the medium, which are assumed to be independent of polarization and therefore scalar. In what follows, calligraphic symbols will be used to denote 4-component vectors of Stokes components. 

We assume that the medium heterogeneities are purely absorbing, that is $\mu_s({\bf r}) = \bar{\mu}_s$ and $\mu_t({\bf r}) = \bar{\mu}_s + \bar{\mu}_a + \delta\mu_a({\bf r})$. Here $\bar{\mu}_s$ and $\bar{\mu}_a$ are the constant background values of the respective coefficients.

An incident collimated laser beam is described mathematically by the source function ${\mathcal S}({\bf r}, \hat{{\bf s}}) = {\mathcal S}_{\rm in} \delta({\bf r} - {\bf r}_{a}) \delta(\hat{\bf s} -  \hat{\bf s}_a)$, where ${\mathcal S}_{\rm in}$ is the Stokes vector for the incident beam. The solution in a homogeneous medium (that is, in a medium with $\delta\mu_a({\bf r}) = 0$), ${\mathcal I}_0 ({\bf r}, \hat{\bf s})$, can be written in the form 
\begin{align}
{\mathcal I}_0({\bf r}, \hat{\bf s}) = G({\bf r}, \hat{\bf s}; {\bf r}_a, \hat{\bf s}_a) {\mathcal S}_{\rm in} \ ,
\end{align}
\noindent
where the Green's function $G({\bf r}, \hat{\bf s}; {\bf r}_{a}, \hat{\bf s}_a)$ is a $4 \times 4$ matrix. Within the validity of the first Born approximation, the solution in a heterogeneous medium, evaluated at the location and in the collimation direction of the detector, can be written as
\noindent
\begin{align}
{\mathcal I}({\bf r}_b, \hat{\bf s}_b) & = {\mathcal I}_0({\bf r}_b, \hat{\bf s}_b) \nonumber \\
& - \int G({\bf r}_b, \hat{\bf s}_b; {\bf r}, \hat{\bf s}) \delta \mu_a({\bf r}) 
G({\bf r}, \hat{\bf s}; {\bf r}_a, \hat{\bf s}_a) {\mathcal S}_{\rm in} d^2 s d^3 r \ .
\label{eqn:pert}
\end{align}
\noindent
This equation shows that the presence of absorptive heterogeneities will result in a detectable variation of the measured specific intensity, in fact, all four Stoks components thereof. We define the data function as the shadow created by the heterogeneities projected onto a given polarization state ${\mathcal S}_{\rm out}$, viz,
\begin{align}
\varPhi &({\bf r}_b, \hat{{\bf s}}_b; {\bf r}_a, \hat{\bf s}_a) \nonumber \\
&= \left. {\mathcal S}_{\rm out} \cdot \Big{[} {\mathcal I}_0 ({\bf r}_b, \hat{{\bf s}}_{b}) - {\mathcal I}({\bf r}_b, \hat{\bf s}_b) \Big{]} \right\vert_{{\mathcal S}({\bf r}, \hat{{\bf s}}) = S_{\rm in} \delta({\bf r} - {\bf r}_a) \delta(\hat{\bf s} -  \hat{\bf s}_a) } \ .
\end{align}
\noindent
Here the dot product of two Stokes vectors is evaluated according to the usual rules, that is, ${\mathcal I}_1 \cdot {\mathcal I}_2 = I_1 I_2 + Q_1 Q_2 + U_1 U_2 + V_1 V_2$. Experimentally, projection onto the polarization state ${\mathcal S}_{\rm out}$ is achieved by using an appropriate polarization filter in front of the detector and may involve a subtraction of two different measurements. Note also that acquisition of the data function $\varPhi$ requires either a differential measurement involving the heterogeneous and a reference (homogeneous) medium or an analytical expression for $G$. Assuming that $\varPhi$ has been measured, we can relate it to the medium heterogeneities through the linear integral equation of the form
\begin{align}
\varPhi({\bf r}_b, \hat{{\bf s}}_b; {\bf r}_a, \hat{\bf s}_a)  = \int \left[ {\mathcal S}_{\rm out} \cdot K({\bf r}_b, \hat{\bf s}_b, {\bf r}_a, \hat{\bf s}_a; {\bf r})
{\mathcal S}_{\rm in}\right] \delta \mu_a({\bf r}) d^3 r \ ,
\label{eqn:sens_integral}
\end{align}
\noindent
where 
\begin{align}
\label{K_def}
K({\bf r}_b, \hat{\bf s}_b, {\bf r}_a, \hat{\bf s}_a; {\bf r}) =
\int G({\bf r}_b, \hat{\bf s}_b; {\bf r}, \hat{\bf s}) G({\bf r}, \hat{\bf s}; {\bf r}_a, \hat{\bf s}_a) d^2 s \ .
\end{align}
The $4\times 4$ matrix $K$ is the sensitivity kernel for vRTE. It is a generalization of a similar scalar kernel that is applicable to unpolarized light~\cite{schotland_07_1}. The additional degrees of freedom in $K$ are associated with using different linearly-independent polarization filters in front of the source and the detector. It is worth noting that, in the scalar case, the sensitivity function can only be positive, as the addition of an absorber at some location can only reduce the measured intensity. While the same is true for the matrix element $K_{11}$ of the sensitivity kernel, other elements are not restricted to be positive due to the fact that the Stokes components $Q$, $U$, $V$ can change sign. Also, the element $K_{11}$ is expected to be close but not identical to the sensitivity kernel of the scalar RTE, which was considered, for example, in~\cite{schotland_07_1}. 

Just like in the scalar case, the definition (\ref{K_def}) involves an angular integral of two Green's functions. One of these functions gives the specific intensity in the reference medium due to the source and is represented in Fig.~\ref{fig:coords} by the solid line leaving the source and arriving at ${\bf r}$. The other function can be interpreted as the specific intensity due to an internal source and is represented by the solid line leaving the volume element at ${\bf r}$ and arriving at the detector. 

We are interested in computing $K$ by MC simulations. However, direct application of (\ref{K_def}) requires computing a new Green's function for every interior point of the medium, e.g., for each voxel if the problem is discretized. Of course, if the medium is an infinite slab, we can use translational invariance to reduce the number of required computations dramatically. In what follows, we utilize this approach as well as certain reciprocity relations for the phase matrix $Z$ to show that only one or few MC simulations are required to compute the kernel $K$.

We now proceed with deriving the reciprocity relation. We note that, in macroscopically isotropic media, the phase matrix satisfies the reciprocity relation~\cite{hovenier_69_1}
\begin{align}
\label{zet}
Z(-\hat{\bf s}^\prime, -\hat{\bf s}) = P Z^T(\hat{\bf s}, \hat{\bf s}^\prime) P \ ,
\end{align} 
\noindent
where $P = \textup{diag}[1,1,-1,1]$. Note that \eqref{zet} is a generalization of the relation $A(-\hat{\bf s}^\prime, -\hat{\bf s}) = A(\hat{\bf s}, \hat{\bf s}^\prime)$ for the phase function $A$ of the scalar RTE. Now, consider the scattering-order expansion of the Green's function $G({\bf r}_{\rm out}, \hat{\bf s}_{\rm out}; {\bf r}_{\rm in}, \hat{\bf s}_{\rm in})$, where ${\bf r}_{\rm in}$ and ${\bf r}_{\rm out}$ are two generic points inside the medium or on its boundary. The expansion can be written as a sum of terms involving $n$ scattering events , and each of these terms is an integral over a set of ``internal'' positions and directions. This set of internal variables defines a photon path -- a piece-wise linear trajectory connecting ${\bf r}_{\rm in}$ to ${\bf r}_{\rm out}$. Since ballistic propagation between two scattering vertices does not change the state of polarization, the Mueller matrix of a photon that has traveled along a given path involving $n$ vertices ($n=1,2,\ldots$) is of the form
\begin{equation}
\label{M_n_forward}
M_{\rm forward} = Z(\hat{\bf s}_{\rm out}, \hat{\bf s}_n) Z(\hat{\bf s}_n, \hat{\bf s}_{n-1}) \ldots Z(\hat{\bf s}_1, \hat{\bf s}_{\rm in}) \ . 
\end{equation}
\noindent
The above expression contains a product of $n+1$ phase matrices. Note that not every set of vertices ${\bf r}_{\rm in}$, ${\bf r}_1$, ${\bf r}_2$, \ldots , ${\bf r}_{\rm out}$ and directions $\hat{\bf s}_{\rm in}$, $\hat{\bf s}_{\rm 1}$, \ldots, $\hat{\bf s}_{\rm out}$ defines a path. However, if a path can be defined, it is unique. We will restrict attention to the set of variables that define a path. Only such sets of vertices and directions contribute to the scattering order expansion of the Green's function. Note also that the vertex positions are needed to define a path but do not enter the expression (\ref{M_n_forward}). What is important for us is that, for each direct path, there also exists a reverse path whose Mueller matrix is
\begin{align}
\label{M_n_backward}
M_{\rm backward} = Z(-\hat{\bf s}_{\rm in}, -\hat{\bf s}_1) Z(-\hat{\bf s}_1, -\hat{\bf s}_2) \ldots Z(-\hat{\bf s}_n, -\hat{\bf s}_{\rm out}) \ . 
\end{align}
\noindent
It can be seen that, if (\ref{zet}) holds, then $M_{\rm backward} = PM_{\rm forward}^T P$. Since the Green's function $G({\bf r}_{\rm out}, \hat{\bf s}_{\rm out}; {\bf r}_{\rm in}, \hat{\bf s}_{\rm in})$ is a linear superposition of various terms of the form (\ref{M_n_forward}) while $G({\bf r}_{\rm in}, -\hat{\bf s}_{\rm in}; {\bf r}_{\rm out}, -\hat{\bf s}_{\rm out})$ is a superposition of the terms (\ref{M_n_backward}) (with exactly the same weights), we have derived the reciprocity relation
\begin{align}
\label{reciprocity}
G({\bf r}_{\rm in}, -\hat{\bf s}_{\rm in}; {\bf r}_{\rm out}, -\hat{\bf s}_{\rm out}) = P G^T({\bf r}_{\rm out}, \hat{\bf s}_{\rm out}; {\bf r}_{\rm in}, \hat{\bf s}_{\rm in}) P \ .
\end{align}
This is the main theoretical result of this letter. Although it follows straightforwardly from the known relation (\ref{zet}), we are not aware of any prior publications in which this reciprocity relation is explicitly stated. 

Now we show how the reciprocity relation can be used to simplify the computation of the sensitivity kernel $K$. Namely, we set  ${\bf r}_{\rm in} = {\bf r}$, $\hat{\bf s}_{\rm in} = \hat{\bf s}$ and ${\bf r}_{\rm out} = {\bf r}_b$, $\hat{\bf s}_{\rm out} = \hat{\bf s}_b$ in (\ref{reciprocity}) (see Fig.~\ref{fig:coords} for a illustration of the relevant geometry) and obtain
\begin{align}
\label{Gb_Ga}
G({\bf r}_b, \hat{\bf s}_b; {\bf r}, \hat{\bf s}) = P G^T({\bf r}, -\hat{\bf s}; {\bf r}_b, -\hat{\bf s}_b) P \;.
\end{align}
\noindent
The resultant simplification is especially significant if $\hat{\bf s}_b = - \hat{\bf s}_a$, as is shown in the figure. We then use the translational invariance of the Green's function 
to write
\begin{align}
\label{Gb_Ga_1}
G({\bf r}_b, \hat{\bf s}_b; {\bf r}, \hat{\bf s}) = P G^T({\bf r} + {\bf r}_{ab}, -\hat{\bf s}; {\bf r}_a, \hat{\bf s}_a) P \ ,  \ \ {\rm if} \ \hat{\bf s}_b = - \hat{\bf s}_a \ ,
\end{align}
\noindent
where ${\bf r}_{ab} = {\bf r}_a - {\bf r}_b$. Thus, the sensitivity kernel $K$ can now be expressed as
\begin{align}
\label{K_def_1}
K({\bf r}_b, \hat{\bf s}_b, {\bf r}_a, \hat{\bf s}_a; {\bf r}) &= \int P G^T({\bf r} + {\bf r}_{ab}, -\hat{\bf s}; {\bf r}_a, \hat{\bf s}_a) P \nonumber \\
& \times G({\bf r}, \hat{\bf s}; {\bf r}_a, \hat{\bf s}_a) d^2 s \ ,  \ \ {\rm if} \ \hat{\bf s}_b = - \hat{\bf s}_a \ .
\end{align}
\noindent
The important point here is that the above expression involves only one Green's function of the generic form $G({\bf r}, \hat{\bf s}; {\bf r}_a, \hat{\bf s}_a)$. This function can be computed by only four independent MC simulations (see below), with a starting point ${\bf r}_a$ and the initial collimation direction $\hat{\bf s}_a$. This Green's function relates an incident arbitrarily polarized collimated beam to the vector specific intensity, defined in the meridian plane, for each position ${\bf r}$ inside the sample and for each direction $\hat{\bf s}$. Computing this function numerically requires keeping track not only of the voxels visited by a photon (and its polarization state arriving at the voxel), but also of its incoming direction. It is not very typical for MC simulations to keep track of the incoming directions in the photon history. Definitely, this requires a larger statistical sample and, in addition, defining some sort of discrete ordinates, which can be numerically problematic. However, if the separation between the source and the detector $r_{ab}$ is sufficiently large, the integration in (\ref{K_def_1}) takes place in the spatial regions where the angular dependence of at least one of the Green's functions involved is relatively weak. We therefore can adopt the following approach to computing the angular dependence of the Green's function.

Firstly, a given MC simulation produces the vector specific intensity ${\mathcal I}({\bf r}, \hat{\bf s})$ for a given polarization of the source, ${\mathcal S}_{\rm in}$. The $4\times 4$ matrix of the Green's function is then obtained by repeating the MC process for four linearly-independent and physically-realizable incident states of polarization. We will expand each component of ${\mathcal I}({\bf r}, \hat{\bf s})$ in spherical functions $Y_{lm} (\hat{\bf s})$, viz,
\begin{align}
\label{I_i}
I({\bf r}, \hat{{\bf s}}) = \sum_{l=0}^{l_{\rm max}}\sum_{m=-l}^{l} i_{lm} ({\bf r}) Y_{lm} (\hat{\bf s}) \ ,
\end{align}
\noindent
and similarly for the $Q$, $U$, and $V$ components. Here $l_{\rm max}$ is the truncation order and the functions $i_{lm}({\bf r})$, $q_{lm}({\bf r})$, $u_{lm}({\bf r})$, $v_{lm}({\bf r})$ are to be computed numerically. It can be easily shown that, in a stochastic MC process and for each voxel containing the point ${\bf r}$,  
\begin{equation}
i_{lm}({\bf r})  \xrightarrow[N\rightarrow \infty]{} \frac{1}{N} \sum_{j=1}^N I_j Y_{lm}(\hat{\bf s}_j) \ ,
\end{equation}
\noindent
where $N$ is the total number of photons used in the MC simulation, $\hat{\bf s}_j$ ($j=1,2,\ldots, N$) are the incoming direction of the photon entering the voxel and $I_j$ are the respective first components of the Stokes vector of the incoming ray (defined with respect to the meridian plane, as usual). Similar computational formulas can be written for the remaining three coefficients $q_{lm}$, $u_{lm}$ and $v_{lm}$. Thus, for each voxel and each incident state of polarization, we will compute and store in memory $4(l_{\rm max}+1)^2$ coefficients to represent the angular dependence of the specific intensity. The Green's function can then be calculated by using these coefficients and four different incident polarization states (unpolarized, Q-polarized, U-polarized, and V-polarized), following the process introduced in~\cite{bartel_00_1}. Once the Green's function is obtained, the sensitivity kernel, $K$ can be calculated from Eq.~(\ref{K_def_1}) analytically. Here we can use the relation $Y_{lm}(-\hat{\bf s}) = (-1)^l Y_{lm}(\hat{\bf s})$ and orthogonality of the spherical functions.

\begin{figure}
\begin{center}
\includegraphics[width=88mm]{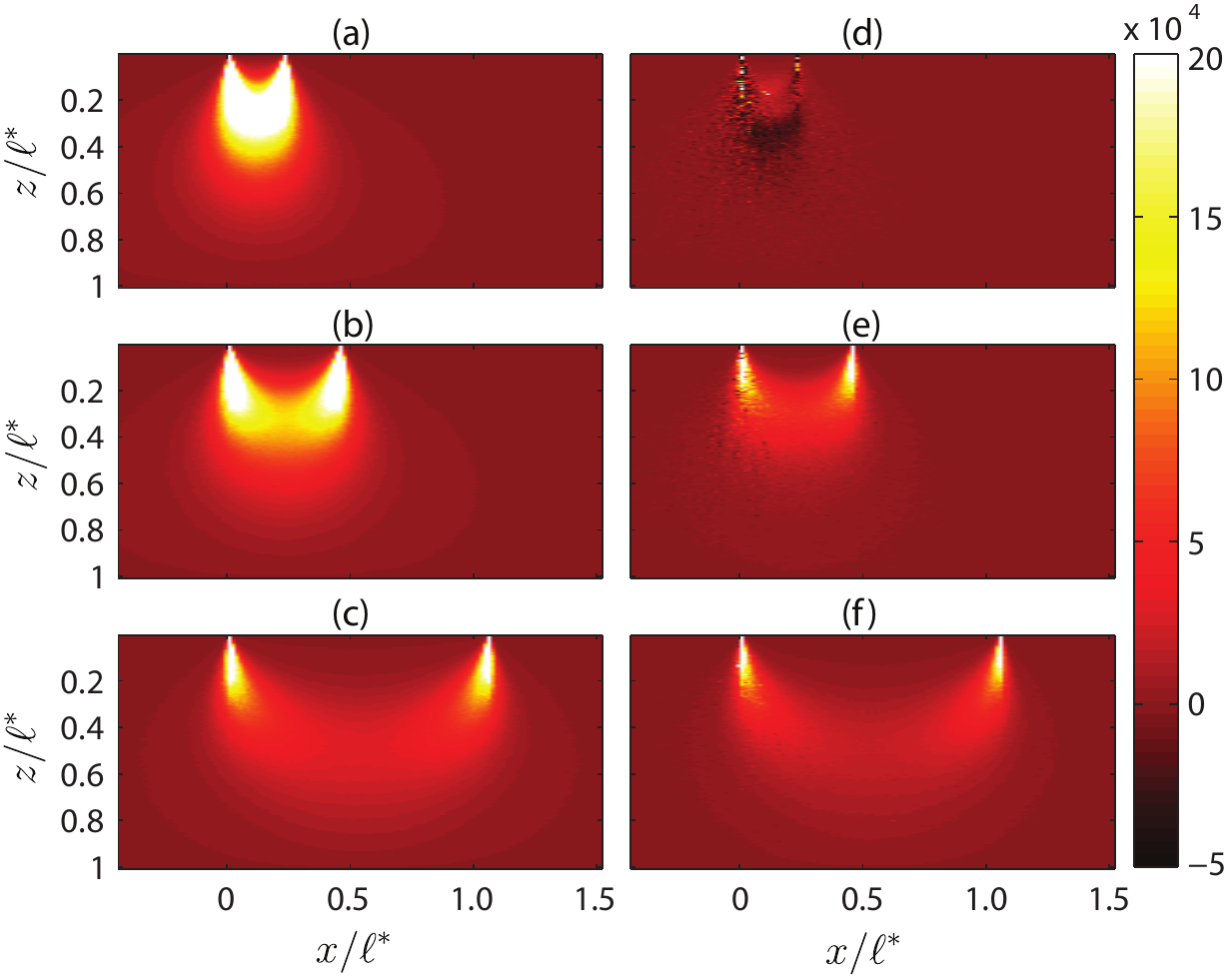}
\caption{Matrix elements of the dimensionless sensitivity kernel $(\ell^*)^2 K$ for normal incidence and normal detection. The matrix element $K_{11}$ is shown in Panels (a-c) and the linear combination  $K_{41} + K_{44}$ is shown in panels (d-f). From top to bottom, the source-detector separation is $0.225\ell^*$, $0.45\ell^*$ and $1.05\ell^*$.}
\label{fig:normal}
\end{center}
\end{figure}

We now show several examples of the computed kernel $K({\bf r}_b, \hat{\bf s}_b, {\bf r}_a, \hat{\bf s}_a; {\bf r})$, more specifically, its various physically accessible matrix elements. In Fig.~\ref{fig:normal}, the sensitivity kernel is shown as a function of ${\bf r}$ for normal illumination and detection. The MC process was implemented in a macroscopically-homogeneous slab with the albedo $\bar{\mu}_s / \bar{\mu}_a = 500/0.03$, which corresponds to the typical optical parameters of soft biological tissues, e.g., $\bar{\mu}_s = 500 {\rm cm}^{-1}$ and $\bar{\mu}_a =0.03{\rm cm}^{-1}$. The actual values of the background optical coefficients is insignificant since all spatial dimensions were scaled by the transport mean free path, $\ell^* = 1/[\bar{\mu}_a + (1-g)\bar{\mu_s}]$ where $g$ is the scattering asymmetry parameter. The phase matrix $Z$ was computed by using Mie theory for spherical inclusions of the the refractive index $n_i$ and radius $a$ in a homogeneous host of refractive index $n_h$. We have taken $n_h=1.33$ (water in the visible spectral range), $n_i = 1.38$ and the size parameter of the inclusions $x = n_h k a = 7.15$, where $k=\omega/c$ is the free space wave number at the working frequency. The scattering asymmetry parameter (the average cosine of the scattering angle) for these scatterers is $g=0.95$. The slab depth was $1\ell^*$ and the lateral dimensions were large enough not to influence the results. The maximum order of spherical functions used in the expansion of angularly-dependent functions was $l_{\rm max}=15$; we have verified that the integral in (\ref{K_def_1}) is well converged for this $l_{\rm max}$ in all cases.

\begin{figure}
\begin{center}
\includegraphics[width=88mm]{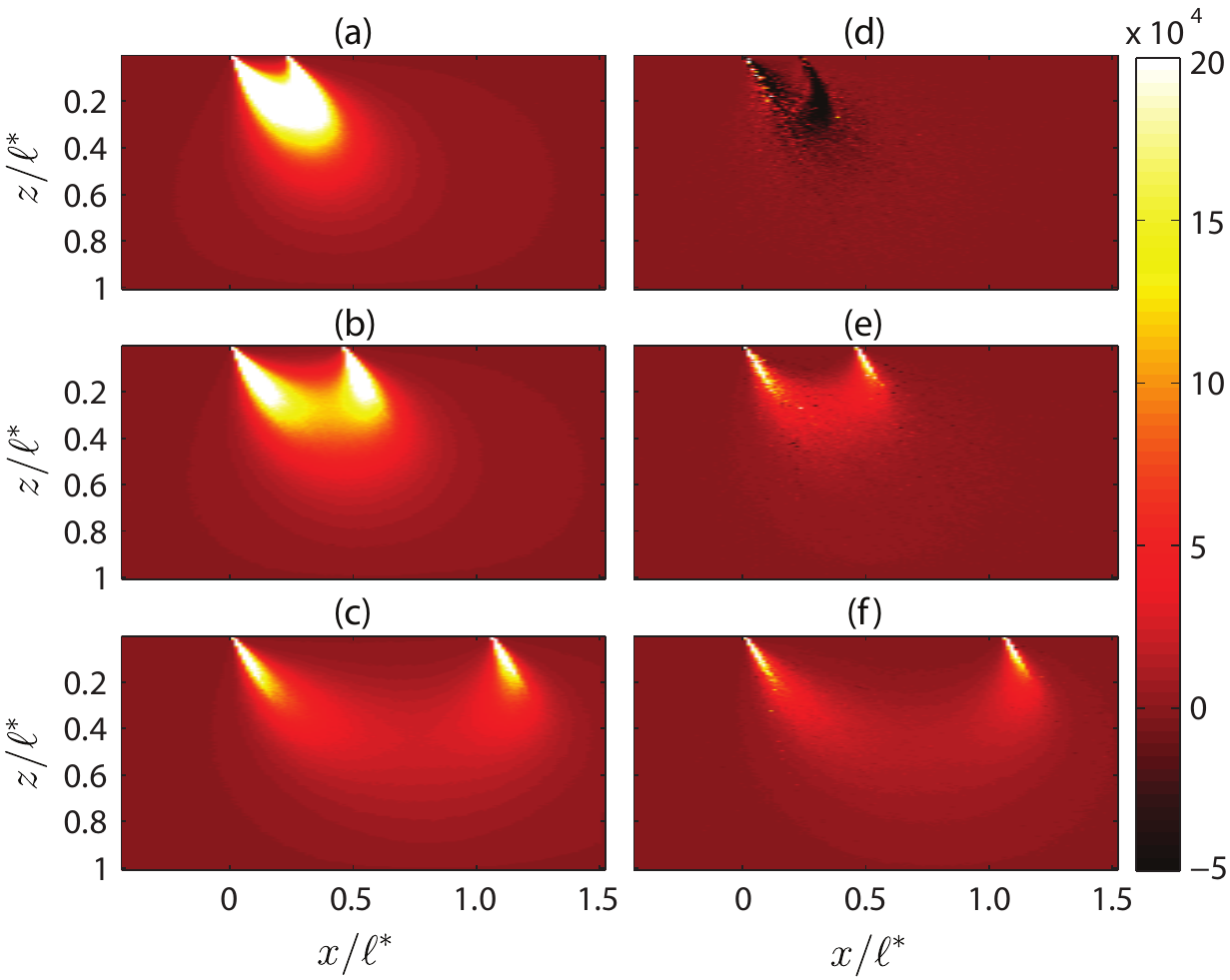}
\caption{Same as in Fig.~\ref{fig:normal} but for off-normal incidence (the incident beam make makes $30^{\circ}$ with the normal).}
\label{fig:angle}
\end{center}
\end{figure} 

In panels (a-c) of Fig.~\ref{fig:normal}, we show the matrix element $K_{11}$ for varying source-detector lateral separation. More specifically, the cross section of $K_{11}({\bf r}_b, \hat{\bf s}_b, {\bf r}_a, \hat{\bf s}_a; {\bf r})$ is shown as a function of ${\bf r}$ in the vertical plane that contains both the source and the detector. We can assume that the variable ${\bf r}$ is restricted so that ${\bf r}=(x,0,z)$ and the cross sections of the medium visualized by the density plots are the $XZ$ planes of the laboratory frame, where the $Z$-axis is perpendicular to the slab. Note that, in the MC simulations, the source position was to the left of the detector. The element $K_{11}$ is relevant if we inject unpolarized light into the medium and performing polarization-insensitive intensity measurements. In this case, ${\mathcal S}_{\rm in} = [1, 0, 0, 0]^T$. It can be seen that the sensitivity kernel alters significantly with the change in source-detector separation. As expected, the ``bridge'' that connects the source to the detector in the intermediate region lowers to greater depths as the separation is increased. On the other hand, the greatest sensitivity is to inhomogeneities placed right in front of the source and detector. Such absorbing inhomogeneities have the potential to block the light completely and the corresponding sensitivity is very high. This is, of course, problematic for practical application of optical tomography. One possible solution is utilization of transparent and homogeneous matching layers (gels) or similar means of excluding the regions of very high sensitivity from the volume in which tomographic reconstructions are sought.

In Panels (d-f) of Fig.~\ref{fig:normal}, we also plot the linear combination $K_{41} + K_{44}$. This matrix element is relevant if we use right-circularly polarized source and measure the Stokes component $V$ on exit. In this case, the polarization state of the incident beam is  ${\mathcal S}_{\rm in} = [1, 0, 0, 1]^T$. We see here that for the two larger source-detector separations, sensitivity kernel is positive-valued at all locations, similar to the case of the $K_{11}$. The most notable difference is that the areas of high sensitivity are reduced in Panels (e-f). The lower sensitivity results from the fundamental inequality for polarized light $I^2 \geq Q^2 + U^2 + V^2$, and the inevitable depolarization of the incident source due to multiple scattering. Also, for the smaller source-detector separation shown in Fig.~\ref{fig:normal}(d), a region of negative sensitivity appears close to the surface, implying that an added absorber at these locations increases the measured Stokes parameter $V$. This counter-intuitive behavior can be explained by considering that certain photon trajectories that penetrate only superficially are more likely to experience a flip in their helicity, i.e., a change of the sign of $V$~\cite{mackintosh_89_1}. Thus, when an absorber removes these photons, the measured signal becomes more positive. This example highlights the complexity of polarization-sensitive measurements. An efficient numerical tool to compute the sensitivity kernel $K$, which we have developed here, allows one to fully take advantage of these unexpected features in image reconstruction.

In Fig.~\ref{fig:angle}, we plot the results of a similar simulation but for an off-normal angles of incidence and detection ($30^{\circ}$ from the normal). One notable feature of Fig.~\ref{fig:angle}(d), is that the negative region of the sensitivity kernel $K_{41} + K_{44}$ becomes more pronounced since the oblique incidence increases the probability of shallow photon paths that tend to reverse the photon helicity. 

In summary, we have presented an approach to MC calculation of the DOT sensitivity kernel for polarized light. The reduction of the computational complexity was obtained by utilizing a reciprocity relation for vRTE, which we have derived in this letter. The numerical results shown above were restricted to the cases when the source and the detector collimation directions $\hat{\bf s}_a$ and $\hat{\bf s}_b$ are anti-parallel. However, this is not a fundamental limitation of our method. Measurement schemes with $\hat{\bf s}_a$ and $\hat{\bf s}_b$ making the same angle with the normal but not anti-parallel can be handled with equal efficiency. The general case when $\hat{\bf s}_a$ and $\hat{\bf s}_b$ make arbitrary angles with the normal can also be accommodated but requires twice the computation time.

This work has been carried out thanks to the support of the A*MIDEX project (No. ANR-11-IDEX-0001-02) funded by the ``Investissements d'Avenir'' French Government program, managed by the French National Research Agency (ANR).

\bibliographystyle{apsrev}
\bibliography{abbrev,master,book}

\end{document}